# Synthesis and Characterization of TiFe$_{0.7-x}$Mn$_{0.3}$V$_x$ (x=0.05, and 0.1) and Ti$_{1-y}$Ta$_y$Fe$_{0.7}$Mn$_{0.3}$ (y=0.2, and 0.4) Nanostructured Metal Hydrides for Low Temperature Applications

N.G. Anagnostou[1], S.S. Makridis[1,2,3*], E. S. Kikkinides[2], C.N. Christodoulou[3] and A.K. Stubos[2]

[1]Department of Mechanical Engineering, University of Western Macedonia, Bakola & Sialvera St., 50100, GR.

[2]Institute of Nuclear Technology and Radiation Protection, NCSR 'Demokritos', Agia Paraskevi, 15310, GR.

[3]Hystore Technolgies LtD, 30, Spyrou Kyprianou, Ergates Industrial Area, Nicosia, 2643 Cyprus

* Corresponding author

Sofoklis S. Makridis
Lecturer
Department of Mechanical Engineering
University of Western Macedonia
Bakola & Sialvera Street
Greece, GR-50100
Telephone: +30 24610 56752 (office), +30 24610 56152 (lab)
Fax: +30 24610 56601
E-mail: smakridis@uowm.gr

**Abstract**

Metal hydrides (MH) are often preferred to absorb and desorb hydrogen at ambient temperature and pressure with a high volumetric density. These hydrogen storage alloys create promising prospects for hydrogen storage and can solve the energetic and environmental issues. In the present research work, the goal of our studies is to find the influence of partial substitution of small amounts of vanadium and tantalum on the hydrogenation properties of $TiFe_{0.7-x}Mn_{0.3}V_x$ (x=0.05, and 0.1) and $Ti_{1-y}Ta_yFe_{0.7}Mn_{0.3}$ (y=0.2, and 0.4) alloys, respectively. The nominal compositions of these materials are $TiFe_{0.65}Mn_{0.3}V_{0.05}$, $TiFe_{0.6}Mn_{0.3}V_{0.1}$, $Ti_{0.8}Ta_{0.2}Fe_{0.7}Mn_{0.3}$, and $Ti_{0.6}Ta_{0.4}Fe_{0.7}Mn_{0.3}$. All samples were synthesized by arc-melting high purity elements under argon atmosphere. The structural and microstructural properties of the samples were studied by using XRD and SEM, respectively, while the corresponding microchemistry was determined by obtaining EDS measurements at specific regions of the samples. Mapping was obtained in order to investigate atomic distribution in microstructure. Moreover, to ensure the associations between the properties and structure, all samples were examined by an optical microscope for accessional characterization. From all these microscopic examinations a variety of photomicrographs were taken with different magnifications. The hydrogenation properties were obtained by using a Magnetic Suspension Balance (Rubotherm). In this equipment, the hydrogen desorption and re-absorption, can be investigated at constant hydrogen pressures in the range of 1 to 20 MPa (flow-through mode). At least 3.43 wt. % of absorbed hydrogen amount was measured while the effect of substitutions was investigated at the same temperature.

**Keywords:** AB, Intermetallic compounds, TiFe based materials, Metal Hydrides, Hydrogen Storage.

# 1. Introduction

In the last decades, many research works devoted to solid hydrogen storage using binary intermetallic compounds. Especially, hydrides based on intermetallic compounds of transition metals like AB type of alloys, are promising materials for light, safe and economical storage of hydrogen [1,2]. The A element is usually a transition metal or a rare earth metal and tends to form a stable hydride. The B element is often a transition metal and forms only unstable hydrides [3]. The main representative of AB alloys group is TiFe (or FeTi) with CsCl-type of structure which is able to form hydrides (TiFeH and $TiFeH_2$ as hydrides) [4,5,6,8]. Among the different AB compounds, TiFe and related alloys have received most attention because of the low cost and high abundance of the raw materials [1,7]. Libowitz was the first who demonstrated reversible AB intermetallic compounds by using ZrNi in 1958 and the first practical AB hydrides were demonstrated with TiFe around 1970 by Reilly and Wiswall at Brookhaven National Laboratory, USA. TiFe alloys and its substitutional modifications remain the best of the AB intermetallic compounds today [8]. These AB-type alloys without any substitution can store up to 1.9 wt.% of hydrogen at 40 $^oC$ [2,5,9]. The TiFe alloys stores only small ammounts of hydrogen without any activation treatment but requires a high temperature over 400-450 $^oC$ [5,9].

However, an activation treatment is required in order for TiFe to absorb hydrogen at room temperature. Difficulties with activation make TiFe to remain unpractical in spite of their practical features [10,11]. Many research works dedicated to TiFe activation and other substitutions of this alloy base []. To improve the activation of TiFe several approaches have been adopted [10]. Pure titanium generates too stable hydrides with the limiting composition $TiH_2$ [13]. For this fact, several substitutions of Fe by a transition metal like manganese (Mn), vanadium (V) help the formation of secondary phase and decrease the stability of hydrides. These substitutions are used to improve activation properties of TiFe alloys by changing compositions stoichiometry [1,11,12,13]. For example, excess of titanium (Ti) in TiFe, i.e., $Ti_{1+x}Fe$, enables the alloy to be hydride without the activation treatment [1,11]. Hydrogenation and dehydrogenation kinetics are quite sensitive to material preparations, gas purity, system design etc. that it questionable whether any of the compiled rate constants have a fundamental significance [14]. These explain why surfaces get easily poisoned in air and get prevented from hydriding [10]. These types of alloys can be synthesized by using different methods. Arcmelting and mechanochemical synthesis are the most common laboratory widely used methods for these alloys [5].

TiFe-based alloys could be used for several applications of hydride devices like : hydrogen accumulators, thermal sorption compressors, hydride refrigerators and motor vehicles by using waste heat [15,16]. In summary these alloys could be promising solutions for many mobile and stationary applications [17,18].

## 2. Experimental details

All samples were synthesized by arc-melting high purity granules of titanium, iron, manganese, vanadium, tantalum (Ti with purity 99.99%, Fe with purity 99.98%, Mn with purity 99.99%, V with purity 99.9%, Ta with purity 99.9%). The arc-melting procedure took place in a small electric arc furnace with a water-cooled copper hearth with four cavities (for each sample severally) under pure argon atmosphere (99.999 %). It should be noted that the quantities of the pure raw materials which have been used, conform with the stoichiometric proportion. Concisely, one group of four intermetallic compounds (four samples in total) were prepared in once. These four ingots were remelted four times and turned over each time to ensure consistency of the melt process and a good homogeneity to our specimens. The weight of each sample was 3 gr.

After the completion of the arc-melting process, the bulk ingots have been cleaned in order to remove the poisoned surface of oxide films. The milling process was performed with a high energy planetary ball miller (RETSCH PM 400/2) to produce microsized powder from the clear bulk ingots using stainless steel balls and griding jars at weight ratio of ball to metal powders of 10:1. The milling time-varying was 3 to 5 hours because of the brittleness of our compounds for total shattering. The planetary rotation speed was 250 rpm with reversing control under an argon atmosphere. The main purpose of this procedure is that most of hydriding-dehydriding experiments are performed using powders with a distribution of sizes and shapes [14]. Planetary milling has proved to be an effective method for synthesis and modification of the intermetallic hydrogen storage compounds [5].

The mechanically alloyed powders were microstructurally characterized using a Bruker D8 Advance powder diffractometer (X-Ray, Cu-Kα radiation, Bragg Brentano geometry, $2\theta$ range 30–100$^o$, step size 0.02$^o$) at room temperature. Powder X-ray diffraction (PXRD) data were collected from Diffrac Plus XRD Commander. All the patterns were analysed and refined with the Rietveld method using the program FullProf.

Small pieces of each sample were prepared properly after all materialographic preparation requirements (grinding, lapping and polishing) for an optical analysis. All these processes carried out by a Struers LaboPol-5 with a number of discs, depending on the number and type of preparation steps required. The metallographic analysis was conducted with a Leica DM2500 M optical microscope. The aim of this investigation was the microstructure analysis and characterization of our specimens from a big variety of photomicrographs, from different objective lenses (5x, 10x, 20x, 50x, 100x ). All these photomicrographs, were taken by a Leica DFC295 digital camera connected to the optical microscope.

The SEM/EDAX analysis was performed on finely polished and conductive specimens using a JEOL-840A Scanning Electron Microscope with an OXFORD ISIS 300 EDS detector. The standard operating voltage for EDS analysis on the Oxford was 20 kV and the beam current was being kept at – 6000 nA. Moreover, specimen surfaces were scanned with an electron beam, and the reflected (or back-scattered) beam for microchemistry and elementary mapping analysis. The magnification of this procedure fluctuated from 1000 to 4000.

The pressure-composition-temperature (PCT) curves were measured with a Sievert's type apparatus by using a Magnetic Suspension Balance (Rubotherm). In this latter equipment, hydrogen desorption and re-absorption, can be investigated at constant hydrogen pressures in the range from 1 to 20 MPa (flow-through mode with 99.9999% hydrogen) under desirable temperature. Activations were performed at 250 °C under 3.5 MPa of hydrogen. In total, twelve cycles of hydrogenation/dehydrogenation carried out on our specimens under 50 °C for 9 h.

## 3. Experimental results and discussion

As shown in Fig.1. and Table 1, Rietveld analysis has been performed on all samples. Through this analysis we observed that the main phase of the first three samples is the hexagonal C14 (Laves phase type $MgZn_2$) and the hexagonal C36 (Laves phase type $MgNi_2$) for the last one. Maybe, the last specimen is a minor exception as the tantalum substitutes titanium, gives a different phase in comparison with the other samples. The main phase of the alloys is the Ti-based solid solution phase with a HCP structure (Table 2) which corresponds to P36/mmc space group, like these four specimens. Obviously, titanium quantities of each sample play an important role to the formation of phases. These alloys, belong to the same space group (P36/mmc) like most of AB alloys (Table 1). But the main structure of our intermetallic compounds which is HCP comes in contrast with the usual CsCl type of structure of our TiFe basis [10,19]. As we can discern ( from Table 1) the phases change accordingly with the stoichiometry of our specimens. The following table summarizes important information and parameters about the crystal structure of our components.

From a microstructural standpoint, we are able to use these photomicrographs to observe the microstructure and various imperfections of our samples. In the first one (Fig.2.a.) we have the opportunity to discern distinctive traces of nucleation-the formation of very small (often submicroscopic-like 5.946 μm) particles of a new phase [12]. In the second photomicrograph (Fig.2.b.) we can observe something similar to the first but with a lesser extent. Favorable positions for the formation of these nuclei are imperfection sites, especially grain boundaries [20]. On the last two (Fig.2. c. and d.) photomicrographs of sample surfaces we discern many imperfections (like pollutants and oxides - 4.898µm and 6.436 µm) and various concavities from griding and polishing procedure.

These electron micrographs gives us the ability to investigate microchemistry through EDS analysis of specific regions. As we can observe above (Fig.3.) through this consideration, the darkest areas represent regions rich in titanium, one of the basic componets of our alloy basis. The brightest regions are rich in tantalum (like Spectrum 5 at Fig.3.d. with 45.59 wt.%). Besides, the brightness of regions increases in parallel with the content of iron, the other basic element of our alloys basis. On the last sample Spectrum 1 showed increased content of Ti (84.82 wt. %), affecting mainly Mn, compared to another region of the same image (Spectrum 2) where Ti returns at reasonable levels (36.08 wt %). This distinction shows that brighter areas are rich in heavier elements (e.g. Mn). Similar affections of contents we are able to observe through all of our measurements (Table 3 and 4).

Concisely, mapping analysis gives as the faculty to observe element orientations and allocations using the abundance of an element as the intensity of the image (Fig.4,5,6,7.). Clearly, we are able to discern that the brighter regions are richer of each element respectively. Mapping was obtained in order to investigate atomic distribution in microstructure and extra identifying. In the following mapping photomicrographs the darkest areas again represent high contents of titanium. Also the brighter regions of the first sample represents mainly high of manganese and iron respectively. In the analysis of the chemical composition, vanadium was observed in small amount in all phases for each sample which contains vanadium.

From the mapping analysis of Fig.5. vanadium has a uniform distribution on our specimen surface regardless its low content. The last two mapping micrographs are almost alike with an exception of some greater contents of tantalum in the second one accordingly to the stoichiometry (Fig.6. and Fig.7.).

The P-C-T curves (Fig.8.) from our samples consists of three main phases. The first stage of our sample preparation is a cleaning procedure under high pressure of hydrogen (3.0 MPa) and high temperature (about 250 $^{o}$C). This step is necessary for the proper preparation of our specimens. During this operation, hydrogen reacts with various pollutants of our samples and get removed from the system. This technique allows material surfaces to get cleaned from impurities such as oxygen, oxides and moisture, elements and compounds, which we can meet under ambient conditions. The reaction of impurities with hydrogen requires energy, for this reason the temperature is kept high. Moreover, abrupt decreases of pressure and hydrogen content, means system evacuations. This cleaning procedure is repeated at least twice and the total duration is about 4 hr.

The second and optional stage of material preparation, are the activation cycles. The pressure at this point takes the highest value of 3.5 MPa and remains constant (as flow-throug mode) while temperature reduces with a constant rate until 50 $^{o}$C. Since the reaction of hydride formation is exothermic process [21], is expected as the temperature decreases, the material absorbs hydrogen, thus increasing the hydrogen content [wt.%] of the specimen. The maximum values of hydrogen content at this stage of measurement ranges from 3.1 up to 3.4 wt.% hydrogen (Fig.8.). This process helps these metal hydrides to form clear and active surfaces. Through these cycles metal grains break into smaller and finally hydrogen atoms easily find their interstitials positions by absorption. This entire cleaning-activating procedure is repeated at least twice and the total duration is about 7 hr for each specimen. All samples could be activated under these conditions. A large number of works devoted to hydrogen absorption and activation properties of TiFe alloys [22, 23]. An important factor for all these progresses which should be cited is the purity of hydrogen. If the hydrogen we use for our measurements is not enough pure then we will have to deal with surface poisoning. This phenomenon is very usual to these procedures as we have to face surface and hydrogen contaminants. Impure hydrogen (with $O_2$, $CO_2$ for impurities) and surface poisoning could mean a loss of hydrogen absorption and desorption kinetics [23,24].

The last stage of these treatments are hydrogen absorption/desorption measurements. In total, twelve cycles of hydrogenation/dehydrogenation carried out on our specimens under 50 °C (a low temperature) for almost 19 hr (Fig.8.). In the beginning of this phase we vacuum the system in order to remove all hydrogen content by reducing pressure almost to zero and raising temperature ( up to 250 °C). This serves making a "pure", "clear" material without hydrogen for next maximum absorption/desorption measurements. Although, a hydrogen content (about 1wt.% of hydrogen) remains inside to our specimens before the beginning of hydrogen absorption/desorption measurements and attests the creation of a stable hydride which need high temperatures to absorb whole hydrogen contents (800-1000 °C) [25]. These hydrogen contents owing to the stable hydrides and firmly increase when the temperature drops to 50 °C for hydrogen absorption/desorption cycles. These increasing contents of our samples fluctuates between 2.18 - 2.78 wt.% of hydrogen with a steady average increasing slope of 0.007068 wt.%/ hr. Besides, the intense of this phenomenon obstructs the progress of hydrogen desorption. Moreover, from absorption and desorption kinetics we can assume that our alloys behave quite dynamically. This deduction stems from the slope values of our hydrogenation/dehydrogenation measurements. During the hydrogenation-dehydrogenation progress samples absorb hydrogen with average rates of 2.68, 3.98, 4.03 and 2.919 wt. %/ hr and absorb hydrogen with average rates of 4.6, 4.86, 4.53 and 5.017 wt. %/ hr  for each one respectively. The duration of each absorption/desorption cycle demands and takes about 19-20 min. to absorb and about 12 min. to desorb hydrogen while taking into consideration the contents of the stable hydrides. On the Table 5 we can observe the maximum absorptions.  As we can discern (Table 5), the influences of increasing vanadium and tantalum contents to our compositions have the same slight alterations (0.11wt.% for the first two and 0.10wt.% for the last two) to our maximum hydrogen absorptions. A similar observation achieved for $TiFe_{0.8}Mn_{0.1}V_x$ (x = 0, 0.05 and 0.1) where the total absorption capacity increases slightly with the amount of vanadium [9]. The fact is that the last two compositions which contains tantalum and form hexagonal C36 phase achieve the maximum results. Coincidently, it should be cited that the last specimen ($Ti_{0.6}Ta_{0.4}Fe_{0.7}Mn_{0.3}$) was this one which achieved the highest absorption rate as we mentioned before.

## 4. Conclusion

The AB-type hydrogen storage alloys, TiFe$_{0.7-x}$Mn$_{0.3}$V$_x$ (x=0.05, and 0.1) and Ti$_{1-y}$Ta$_y$Fe$_{0.7}$Mn$_{0.3}$ (y=0.2, and 0.4) were synthesized by using an arc-melting. All of our samples are TiFe based alloys and took every suitable preparation for all measurements. The XRD analysis showed that obtained powders are with composite, nanocrystalline-amorphous microstructure. Through Rietveld analysis we observed hexagonal C14 and C36 as main phases. These hexagonal phases come in contrast with the usual CsCl structure of TiFe based alloys. For extra characterization and observation we used optical techniques. Moreover, hydrogenation/dehydrogenation properties was measured and analysed through P-C-T curves. From the first two samples the total absorption capacity increases slightly with higher amounts of vanadium. With all these treatments we discerned that Ti$_{0.8}$Ta$_{0.2}$Fe$_{0.7}$Mn$_{0.3}$ with hexagonal C36 phase and highest content of tantalum reaches the highest hydrogen absorption of 3.62 wt.%. So, the substitution of tantalum increases absorption.

In conclusion, the AB type of hydrides and especially rechargeable TiFe-based alloys could be a promising solution for hydrogen storage applications.

**Acknowledgments**
This work was supported by the ATLAS-H2 European Project with contract number: PIAP-GA-2009-251562.

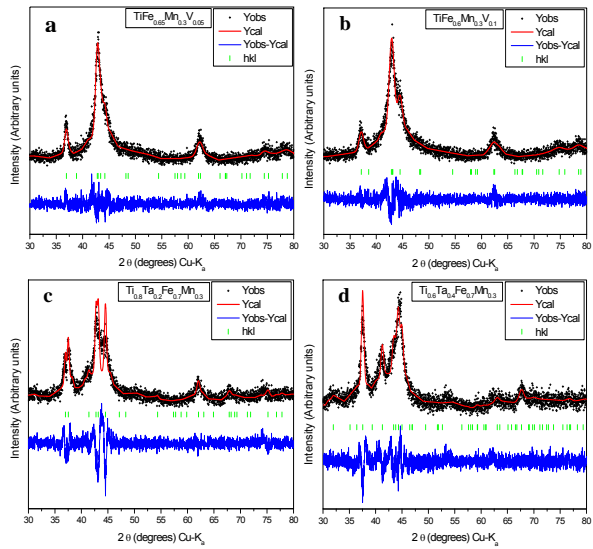

**N.G. Anagnostou, Synthesis And Characterization Of TiFe$_{0.7-x}$Mn$_{0.3}$V$_x$ (x=0.05, and 0.1) And Ti$_{1-y}$Ta$_y$Fe$_{0.7}$Mn$_{0.3}$ (y=0.2, and 0.4) Nanostructured Metal Hydrides For Low Temperature Applications**

**Fig.1.** Powder X-ray diffraction patterns of **a.** TiFe$_{0.65}$Mn$_{0.3}$V$_{0.05}$, **b.** TiFe$_{0.6}$Mn$_{0.3}$V$_{0.1}$, **c.** Ti$_{0.8}$Ta$_{0.2}$Fe$_{0.7}$Mn$_{0.3}$, **d.** Ti$_{0.6}$Ta$_{0.4}$Fe$_{0.7}$Mn$_{0.3}$.

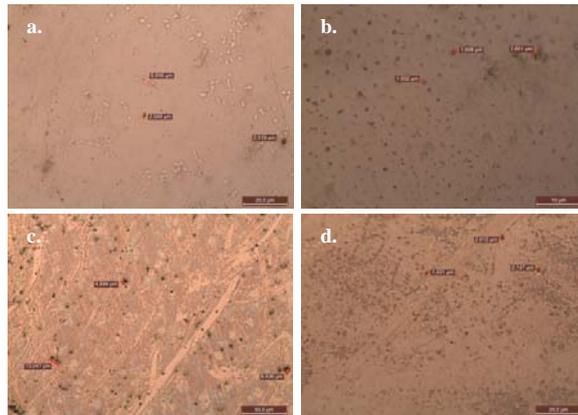

**N.G. Anagnostou, Synthesis And Characterization Of TiFe$_{0.7-x}$Mn$_{0.3}$V$_x$ (x=0.05, and 0.1) And Ti$_{1-y}$Ta$_y$Fe$_{0.7}$Mn$_{0.3}$ (y=0.2, and 0.4) Nanostructured Metal Hydrides For Low Temperature Applications**

**Fig.2.** Photomicrographs from an optical microscope of **a.** TiFe$_{0.65}$Mn$_{0.3}$V$_{0.05}$ with magnification x50, **b.** TiFe$_{0.6}$Mn$_{0.3}$V$_{0.1}$ with magnification x100, **c.** Ti$_{0.8}$Ta$_{0.2}$Fe$_{0.7}$Mn$_{0.3}$ with magnification x20, **d.** Ti$_{0.6}$Ta$_{0.4}$Fe$_{0.7}$Mn$_{0.3}$ with magnification x50.

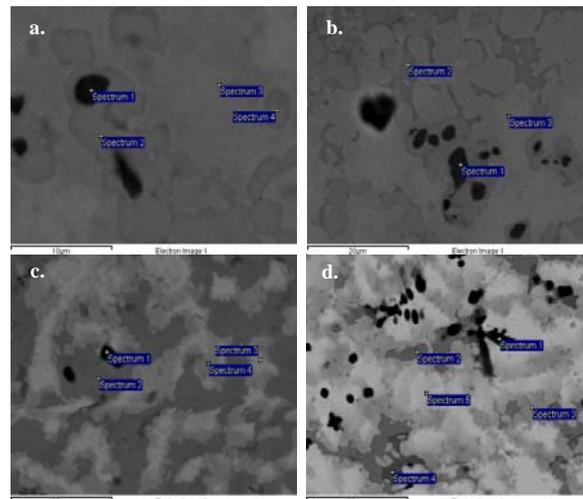

**N.G. Anagnostou, Synthesis And Characterization Of TiFe$_{0.7-x}$Mn$_{0.3}$V$_x$ (x=0.05, and 0.1) And Ti$_{1-y}$Ta$_y$Fe$_{0.7}$Mn$_{0.3}$ (y=0.2, and 0.4) Nanostructured Metal Hydrides For Low Temperature Applications**

**Fig.3.** Scanning electron microscope photomicrographs with EDS analysis spectrums of **a.** TiFe$_{0.65}$Mn$_{0.3}$V$_{0.05}$ with magnification x4000, Backscattered, **b.** TiFe$_{0.6}$Mn$_{0.3}$V$_{0.1}$ with magnification x2000, Backscattered, **c.** Ti$_{0.8}$Ta$_{0.2}$Fe$_{0.7}$Mn$_{0.3}$ with magnification x2000, Backscattered, **d.** Ti$_{0.6}$Ta$_{0.4}$Fe$_{0.7}$Mn$_{0.3}$ with magnification x2000, Backscattered.

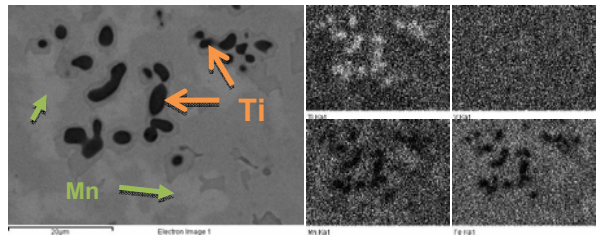

**N.G. Anagnostou, Synthesis And Characterization Of TiFe$_{0.7-x}$Mn$_{0.3}$V$_x$ (x=0.05, and 0.1) And Ti$_{1-y}$Ta$_y$Fe$_{0.7}$Mn$_{0.3}$ (y=0.2, and 0.4) Nanostructured Metal Hydrides For Low Temperature Applications**

**Fig.4.** Backscattered electron micrograph with mapping analysis of TiFe$_{0.65}$Mn$_{0.3}$V$_{0.05}$ with magnification x2000.

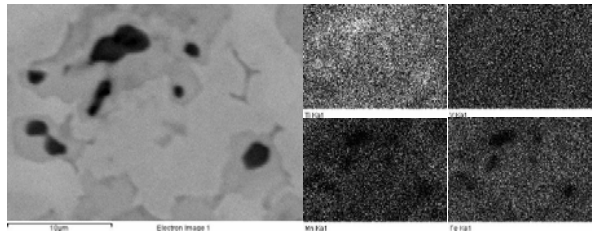

**N.G. Anagnostou, Synthesis And Characterization Of TiFe$_{0.7-x}$Mn$_{0.3}$V$_x$ (x=0.05, and 0.1) And Ti$_{1-y}$Ta$_y$Fe$_{0.7}$Mn$_{0.3}$ (y=0.2, and 0.4) Nanostructured Metal Hydrides For Low Temperature Applications**

**Fig.5.** Backscattered electron micrograph with mapping analysis of TiFe$_{0.6}$Mn$_{0.3}$V$_{0.1}$ with magnification x4000.

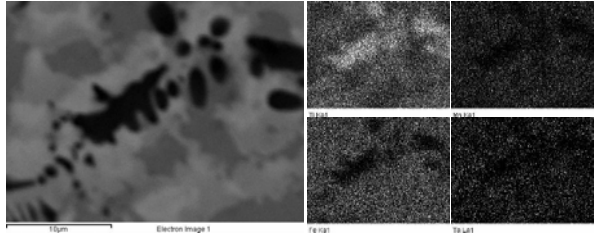

**N.G. Anagnostou, Synthesis And Characterization Of TiFe$_{0.7-x}$Mn$_{0.3}$V$_x$ (x=0.05, and 0.1) And Ti$_{1-y}$Ta$_y$Fe$_{0.7}$Mn$_{0.3}$ (y=0.2, and 0.4) Nanostructured Metal Hydrides For Low Temperature Applications**

**Fig.6.** Backscattered electron micrograph with mapping analysis of Ti$_{0.8}$Ta$_{0.2}$Fe$_{0.7}$Mn$_{0.3}$ with magnification x4000.

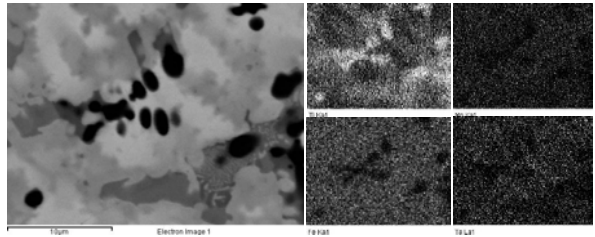

**N.G. Anagnostou, Synthesis And Characterization Of TiFe$_{0.7-x}$Mn$_{0.3}$V$_x$ (x=0.05, and 0.1) And Ti$_{1-y}$Ta$_y$Fe$_{0.7}$Mn$_{0.3}$ (y=0.2, and 0.4) Nanostructured Metal Hydrides For Low Temperature Applications**

**Fig.7.** Backscattered electron micrograph with mapping analysis of Ti$_{0.6}$Ta$_{0.4}$Fe$_{0.7}$Mn$_{0.3}$ with magnification x4000.

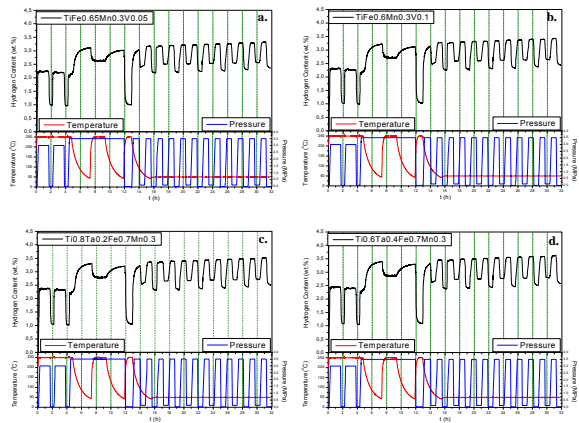

**N.G. Anagnostou, Synthesis And Characterization Of TiFe$_{0.7-x}$Mn$_{0.3}$V$_x$ (x=0.05, and 0.1) And Ti$_{1-y}$Ta$_y$Fe$_{0.7}$Mn$_{0.3}$ (y=0.2, and 0.4) Nanostructured Metal Hydrides For Low Temperature Applications**

**Fig.8.** Pressure-Composition-Temperature curves for **a.**TiFe$_{0.65}$Mn$_{0.3}$V$_{0.05}$, **b.**TiFe$_{0.6}$Mn$_{0.3}$V$_{0.1}$, **c.**Ti$_{0.8}$Ta$_{0.2}$Fe$_{0.7}$Mn$_{0.3}$, **d.**Ti$_{0.6}$Ta$_{0.4}$Fe$_{0.7}$Mn$_{0.3}$.

| Sample | Space Group (No.) | Phase | Unit Cell Parameters (Å) | | Unit Cell Volume (Å$^3$) | Refinement parameters | | |
|---|---|---|---|---|---|---|---|---|
| | | | a | c | | $R_p$ | $R_{wp}$ | $\chi^2$ |
| TiFe$_{0.65}$Mn$_{0.3}$V$_{0.05}$ | P63/mmc | C14 | 4.93145 | 8.42537 | 523.3148 | 76.2 | 53.8 | 1.27 |
| TiFe$_{0.6}$Mn$_{0.3}$V$_{0.1}$ | P63/mmc | C14 | 4.85630 | 8.46167 | 518.4465 | 76.8 | 59.1 | 1.45 |
| Ti$_{0.8}$Ta$_{0.2}$Fe$_{0.7}$Mn$_{0.3}$ | P63/mmc | C14 | 4.70525 | 8.47075 | 487.2375 | 113 | 73.9 | 1.87 |
| Ti$_{0.6}$Ta$_{0.4}$Fe$_{0.7}$Mn$_{0.3}$ | P63/mmc | C36 | 4.95456 | 15.9009 | 1014.159 | 99.5 | 61.5 | 1.42 |

**Table 1.** A summary of results from our Rietveld analysis with the main phases and other information of our samples.

| Element | Space Group | No. | Structure | Unit Cell Parameters | | | | | |
|---|---|---|---|---|---|---|---|---|---|
| | | | | a / Å | b / Å | c / Å | α /° | β/° | γ/° |
| Ti | P36/mmc | 194 | HCP | 2.9508 | 2.9508 | 4.6855 | 90 | 90 | 120 |
| Fe | Im-3m | 229 | BCC | 2.8665 | 2.8665 | 2.8665 | 90 | 90 | 90 |
| Mn | I-43m | 217 | CUBIC | 8.9125 | 8.9125 | 8.9125 | 90 | 90 | 90 |
| V | Im-3m | 229 | BCC | 3.03 | 3.03 | 3.03 | 90 | 90 | 90 |
| Ta | Im-3m | 229 | BCC | 3.3013 | 3.3013 | 3.3013 | 90 | 90 | 90 |

**Table 2.** Features and parameters of our basic components.

| EDS Analysis | | Spectrum 1 | | Spectrum 2 | | Spectrum 3 | | Spectrum 4 | |
| --- | --- | --- | --- | --- | --- | --- | --- | --- | --- |
| Element | Sample | Wt.% | At.% | Wt.% | At.% | Wt.% | At.% | Wt.% | At.% |
| Ti K | a | 71.93 | 74.76 | 49.95 | 53.57 | 46.40 | 50.01 | 46.14 | 49.73 |
|  | b | 97.77 | 98.05 | 59.81 | 63.02 | 41.21 | 44.56 | - | - |
| V K | a | 1.32 | 1.29 | 2.25 | 2.27 | 2.57 | 2.61 | 2.99 | 3.03 |
|  | b | 0.36 | 0.34 | 5.34 | 5.29 | 7.21 | 7.33 | - | - |
| Mn K | a | 7.58 | 6.87 | 13.25 | 12.39 | 13.55 | 12.73 | 14.19 | 13.34 |
|  | b | 0.65 | 0.57 | 13.28 | 12.20 | 18.51 | 17.44 | - | - |
| Fe K | a | 19.17 | 17.09 | 34.54 | 31.77 | 37.48 | 34.65 | 36.68 | 33.91 |
|  | b | 1.22 | 1.05 | 21.56 | 19.48 | 33.07 | 30.67 | - | - |

**Table 3.** Summary table from EDS analysis measurements for **a.** $TiFe_{0.65}Mn_{0.3}V_{0.05}$ and **b.** $TiFe_{0.6}Mn_{0.3}V_{0.1}$.

| EDS Analysis | | Spectrum 1 | | Spectrum 2 | | Spectrum 3 | | Spectrum 4 | | Spectrum 5 | |
|---|---|---|---|---|---|---|---|---|---|---|---|
| Element | Sample | Wt.% | At.% | Wt.% | At.% | Wt.% | At.% | Wt.% | At.% | Wt.% | At.% |
| Ti K | c | 31.67 | 40.14 | 38.25 | 45.78 | 29.26 | 37.62 | 22.84 | 31.34 | - | - |
|  | d | 84.82 | 90.71 | 36.08 | 44.74 | 27.66 | 38.44 | 56.81 | 66.34 | 12.73 | 20.98 |
| Mn K | c | 14.03 | 15.50 | 11.11 | 11.59 | 14.65 | 16.42 | 14.09 | 16.86 | - | - |
|  | d | 1.85 | 1.72 | 6.70 | 7.24 | 9.57 | 11.59 | 5.58 | 5.69 | 10.46 | 15.02 |
| Fe K | c | 34.76 | 37.79 | 37.46 | 38.46 | 35.26 | 38.88 | 35.51 | 41.79 | - | - |
|  | d | 5.98 | 5.49 | 39.75 | 42.28 | 32.61 | 38.87 | 23.60 | 23.64 | 31.22 | 44.12 |
| Ta M | c | 19.55 | 6.56 | 13.18 | 4.18 | 20.83 | 7.09 | 27.57 | 10.01 | - | - |
|  | d | 7.35 | 2.08 | 17.47 | 5.74 | 30.16 | 11.10 | 14.01 | 4.33 | 45.59 | 19.88 |

**Table 4.** Summary table from EDS analysis measurements for **c.** $Ti_{0.8}Ta_{0.2}Fe_{0.7}Mn_{0.3}$ and **d.** $Ti_{0.6}Ta_{0.4}Fe_{0.7}Mn_{0.3}$.

| | Sample | Maximum Hydrogen Contents [wt.%] |
|---|---|---|
| 1º | $TiFe_{0.65}Mn_{0.3}V_{0.05}$ | 3.32 |
| 2º | $TiFe_{0.6}Mn_{0.3}V_{0.1}$ | 3.43 |
| 3º | $Ti_{0.8}Ta_{0.2}Fe_{0.7}Mn_{0.3}$ | 3.52 |
| 4º | $Ti_{0.6}Ta_{0.4}Fe_{0.7}Mn_{0.3}$ | 3.62 |

**Table 5.** A summary of maximum hydrogen contents from absorption.